\documentclass[fleqn,10pt]{wlscirep}
\usepackage{graphicx,epstopdf,amsmath,amssymb,colordvi,times}
\usepackage{multirow,bm}
\newcommand{\onlinecite}[1]{\hspace{-1 ex} \nocite{#1}\citenum{#1}}

\def\agt{\,\hbox{\lower0.6ex\hbox{$\sim$}\llap{\raise0.6ex\hbox{$>$}}}\,}
\def\alt{\,\hbox{\lower0.6ex\hbox{$\sim$}\llap{\raise0.6ex\hbox{$<$}}}\,}

\title{Large gap electron-hole superfluidity and shape resonances in coupled graphene nanoribbons}

\author[1,*]{M. Zarenia}
\author[2]{A. Perali}
\author[1]{F. M. Peeters}
\author[2]{D. Neilson}
\affil[1]{Department of Physics, University of Antwerp,
Groenenborgerlaan 171, B-2020 Antwerpen, Belgium}
\affil[2]{Dipartamenti di Fisica e di Farmacia, Universit\`{a} di Camerino, 62032 Camerino, Italy}

\affil[*]{mohammad.zarenia@uantwerpen.be}

\begin{abstract}
We predict enhanced electron-hole superfluidity in two coupled electron-hole
armchair-edge
terminated graphene nanoribbons  separated by a thin insulating
barrier.  In contrast to graphene monolayers, the multiple subbands of the nanoribbons are
parabolic at low energy with a
gap between the conduction and valence bands, and with lifted valley degeneracy.
These properties make screening of the electron-hole interaction
much weaker than for coupled electron-hole monolayers, thus boosting the
pairing strength and enhancing the superfluid properties. The pairing strength is further boosted by the quasi-one-dimensional quantum confinement of the carriers, as well as by the large density of states near the bottom of each subband.  The latter magnifies the superfluid shape resonances caused by the quantum confinement.  Several superfluid partial condensates are present
for finite-width nanoribbons with multiple subbands.
We find that superfluidity is predominately in the strongly-coupled BEC and
BCS-BEC crossover regimes,
with large superfluid gaps up to $100$ meV and beyond.
When the gaps exceed the subband spacing, there is significant
mixing of the subbands, a rounding of the shape resonances, and a resulting
reduction in the one-dimensional nature of the system.
\end{abstract}

\begin{document}

\flushbottom
\maketitle
\thispagestyle{empty}

\section*{Introduction}

Superfluidity of spatially separated electrons and holes was predicted nearly
half a century ago \cite{Lozovik} but up to now experimental observation
of this exotic state has been elusive at zero magnetic field, notwithstanding
multiple attempts on very different systems. The discovery of the wonder
material graphene in conjunction with the large band gap insulator hexagonal
boron nitride (h-BN) has raised new hopes for realization of this new
collective many body state.
Recently, superconductivity at temperatures above liquid Helium
has been reported in doped monolayer graphene
by four groups, amplifying  interest in
quantum coherent phenomena in graphene. \cite{SuperGr}

Monolayer graphene is an atomically flat, gapless semiconductor with near
identical conduction and valence bands.
Spatially separated electron-doped and hole-doped
monolayers can be completely insulated from each other with just a few atomic layers of h-BN \cite{Gorbachev2012,Geim2013}.  With such small spatial
separations,
electron-hole pairing by direct Coulomb attraction
would be expected to be strong \cite{Mink,Min2008,comment}.  However
the linear dispersion of the monolayer graphene
energy bands results in very strong Coulomb screening of the electron-hole pairing
attraction, and this suppresses superfluidity in coupled electron-hole graphene monolayers  \cite{Lozovik2012,Gorbachev2012}.
To overcome the strong screening,  Refs.\  \onlinecite{Perali2013}
and \onlinecite{Zarenia2014} proposed using coupled electron-hole
graphene multilayers.  Using multilayers takes advantage of the  nonlinear dispersion
of their energy bands  \cite{McCann,Min2008b},  and the
existence of a  gap between the conduction and valence bands when a gate potential is applied.
\begin{figure}[h]
\centering
\includegraphics[width=8cm]{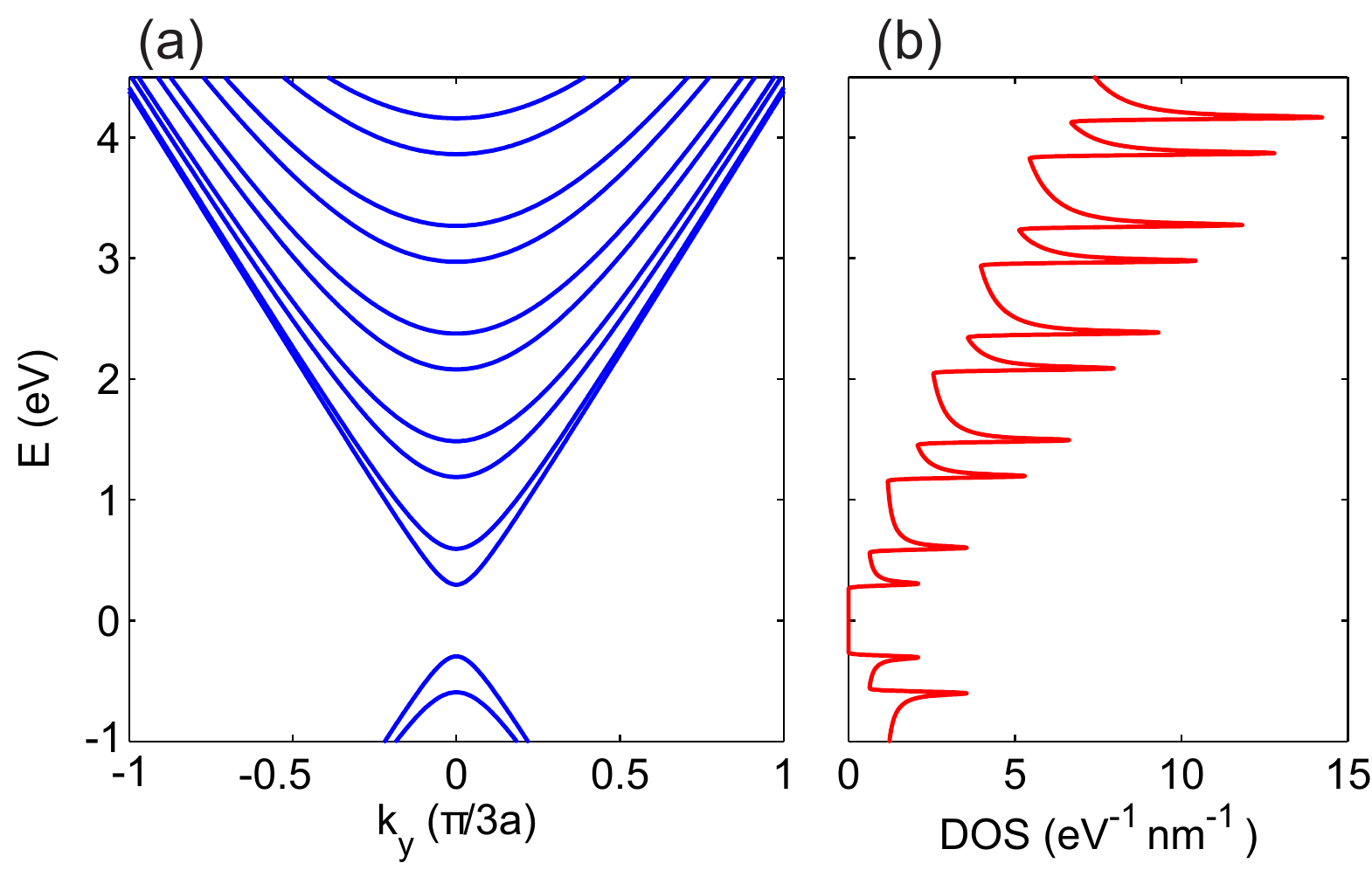}
\caption{(a)  Lowest single-particle energy subbands $\epsilon_j(k_y), j=1,2,\dots$ in an armchair graphene nanoribbon of width $W=2$ nm.
(b) Corresponding density of states DOS$(E)$ in  nanoribbon. Van Hove singularities are visible at  bottom of each subband.}
\label{DOS}
\end{figure}

Here we propose a new design to boost electron-hole pairing and the
onset of superfluidity using nanoribbons etched in monolayer
graphene.
Monolayer sheets of graphene are promising candidates for applications in transparent conductive
films, electronic and opto-electronic devices, actuators, sensors, composites,
and more.  However a serious limitation of graphene monolayers is that field-effect transistor (FET) devices are not possible because the massless nature of the electrons prevents electron
confinement in graphene.
Quasi-one-dimensional graphene nanoribbons with tuneable  band gaps  resolve this  issue, with important implications for the fabrication of novel and ultrafast
electronic nanodevices.   For example, FET devices with 100 GHz switching frequencies
have been fabricated using graphene nanoribbons \cite{Lin2010}.
The nanoribbon edges can be terminated using a variety of different atoms, which opens up application opportunities, in particular for nanoribbons in polymer hosts for
fabrication of novel composite materials \cite{Dimiev2011,Dimiev2013}.
Finally, graphene nanoribbons are showing great promise  as electrode materials for batteries and supercapacitors \cite{Li2013}.

The electronic properties of graphene nanoribbons depend on the type of edge
termination \cite{Kim2007}. We focus on armchair-edge terminated nanoribbons since (i) their subbands
are parabolic around their minima (Fig.\ \ref{DOS}(a)), (ii) there is a sizeable semiconductor-like
energy gap between the conduction
and valence bands, and (iii) the valley degeneracy of monolayer graphene is lifted.
These properties combine to greatly reduce the strength of screening of the  electron-hole pairing
interaction. Note that uniform armchair graphene nanoribbons of
widths  $W\ll 10$ nm have recently been fabricated \cite{Cai}.

\begin{figure}[h]
\centering
\includegraphics[width=8cm]{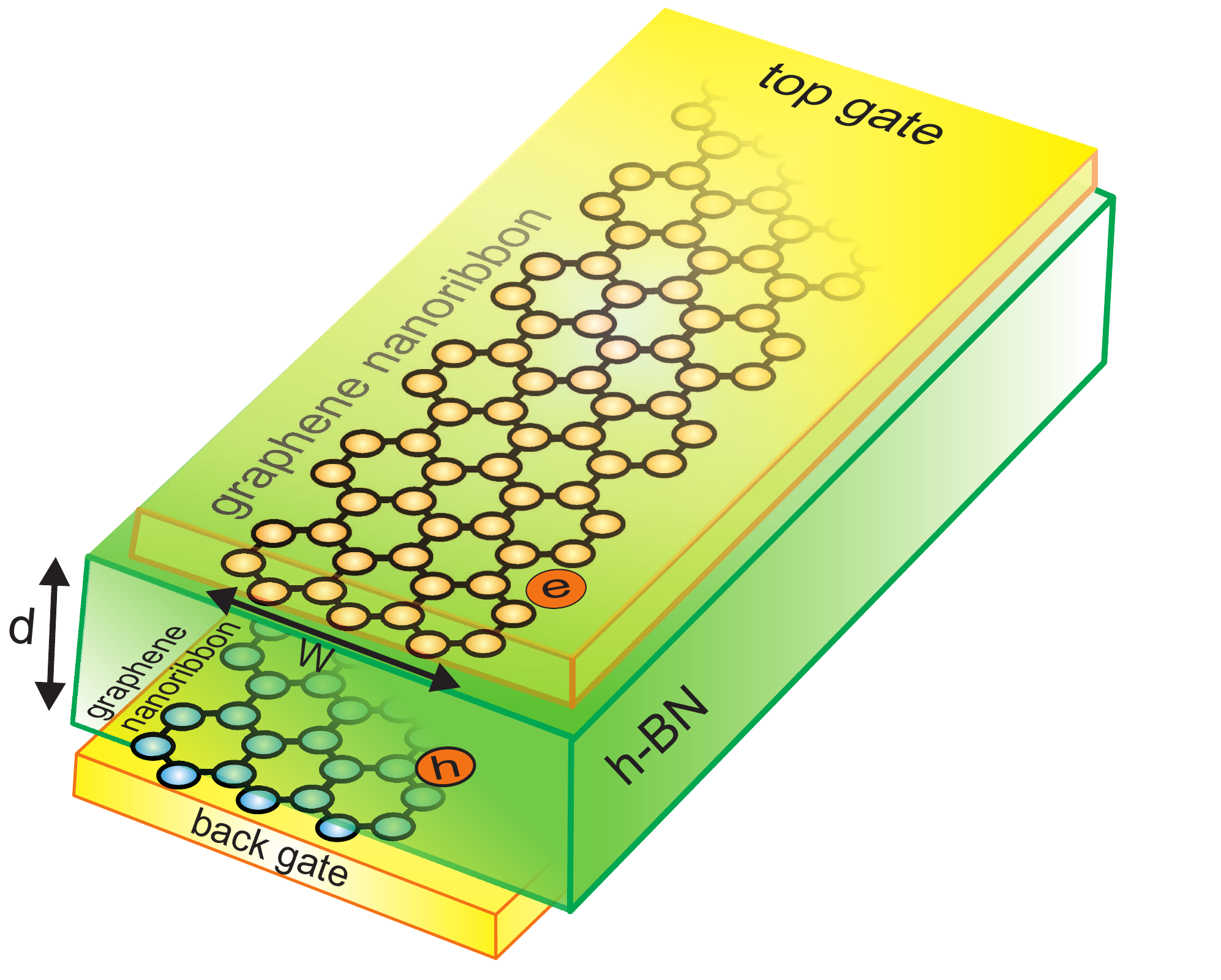}
\caption{Proposed device.  Upper electron-doped and lower hole-doped armchair-edge terminated graphene nanoribbons of  widths $W$ separated by h-BN insulator of thickness $d$. Top and back gates control electron and hole densities.  Gates are separated from  nanoribbons by  h-BN layers. Nanoribbons are independently contacted.
}
\label{schematic}
\end{figure}

Figure  \ref{schematic} shows the device we are proposing.  It consists of two armchair-edge terminated monolayer graphene nanoribbons, one electron-doped and the other hole-doped,    separated by a few atomic layers of a h-BN insulating barrier. The nanoribbons are independently contacted, and
top and back metal gates control the carrier densities.
 \begin{figure}[h]
\centering
\includegraphics[width=9cm]{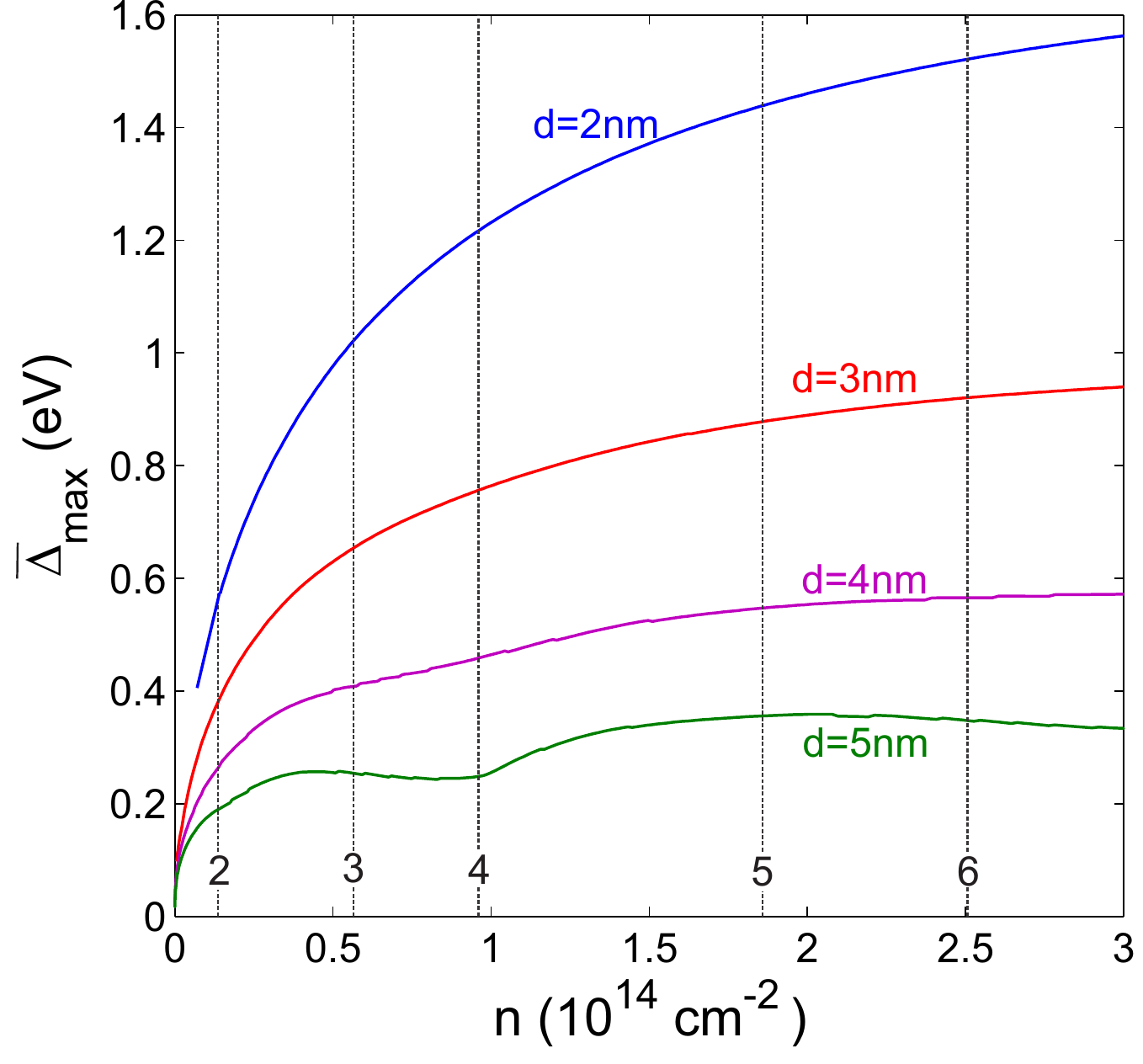}
\caption{Maximum superfluid gap $\overline{\Delta}\,^{\mathrm{max}}$ averaged over the subbands. $d$ is thickness of the insulating barrier separating the nanoribbons.   Nanoribbon width is $W=2$ nm.  Densities at which $E_F$ enters the bottom of a new subband $\epsilon_j$ are indicated by the vertical lines.}
\label{Deltamaxvsd}
\end{figure}

In addition to reducing the effect of screening,  electron-hole pairing
strengths will be further boosted in our proposed system by
the enhanced density of states near the minimum of each subband that arises from the van
Hove singularities of the quasi-one-dimensional nanoribbons, and also by the
quantum confinement of the carriers in the nanoribbons.
Enhancement of superconducting gaps and critical temperature
in striped systems due to shape
resonances and quantum confinement at the nanoscale was
predicted in Refs.\ \onlinecite{Perali1996,Bianconi1997,Bianconi1998}.
Superconductivity has been observed in quasi-one-dimensional systems
including Sn and Al metallic nanowires and  carbon nanotubes,
with enhanced transition temperatures as compared with their bulk values  \cite{Shanenko2006}.

\section*{Methods}
We take the $y$-direction parallel to the nanoribbons, with the carriers
confined in the transverse $x$-direction.  Figure 1(b) shows the
single-particle energy subbands obtained in the continuum model,
$\epsilon_j(k_y)= (\sqrt{3}t a_0/2) \sqrt{k_y^2+k_j^2}$,  \ $j =1,2,\dots$
for an armchair graphene nanoribbon of width
$W=2$ nm.    The intralayer hopping energy $t=2.7$ eV\ \ \cite{Brey2006}
and the graphene lattice constant $a_0=0.24$ nm. $k_j=\left[j\pi/W\right]-\left[4\pi/(3\sqrt{3}a_0)\right]$ is the quantized wave-number
for the $j$-subband in the $x$-direction.  Figure
\ref{DOS}(b)  shows the corresponding density of states DOS$(E)$.
The van Hove singularities coincide with the bottom of each subband.

Not only the  finite width of the nanoribbons but also their multiple occupied subbands make the system only
quasi-one-dimensional.  In addition, in the superfluid state the energy gap mixes close-by subbands.  The quasi-one-dimensionality together with the subband mixing  will
suppress order parameter fluctuations that are responsible for
destroying superfluidity in a pure one-dimensional system.  For these reasons we can
calculate  properties of the superfluid ground state
using mean field theory.

Recently Ref.\ \onlinecite{Abergel} discussed a quasi-condensate of excitons
in coupled electron-hole one-dimensional wires using the weak-coupled BCS
gap equation in the low density limit with only the lowest subband
contributing to the pairing, and with screening neglected.  Since only one
channel was considered, there are no shape resonance effects
at finite densities.  Also, because of the one dimensionality, fluctuations of the order parameter
should be severe  and would strongly suppress
superfluidity.   Interestingly, Ref.\ \onlinecite{Abergel} argues that even in the one-channel case, the finite size of the
nanoribbons would allow for short range superfluid correlations.
In our case the many available channels due to the multiple subbands involved
in the pairing  allow for a  suppression of the critical
fluctuations and it should be straightforward to observe conventional long range superfluidity.

Our calculations are for coupled electron-hole armchair graphene
nanoribbons of equal width $W$ and equal  (two-dimensional) electron and hole densities $n=( r_0W)^{-1}$, where $r_0$ is the average  inter-particle spacing  along the
nanoribbon.   The
subbands $\epsilon_j(k_y),\  j=1,2,\dots$ are identical for the  electrons and holes.

Because of the multiple subband structure, the zero temperature mean field equations for the superfluid state acquire an additional index for the subband $j$.  The equations for the wave-vector dependent superfluid energy
gaps $\Delta_j(k_y) $
for subbands $j$ become,
\begin{equation}
\Delta_j(k_y)  = -\frac{1}{L_y}\sum_{j^\prime}^{j_c}\sum_{k_y^\prime}^{k_c} F_{k_y,j,k_y^\prime,j^\prime}
 V_{\text{e-h}}(k_y-k_y^\prime)  \frac{\Delta_{j^\prime}(k_y^\prime)}{2E_{j^\prime}(k_y^\prime)}\ ,
\label{gapeqn}
\end{equation}
where $L_y$ is the nanoribbon length, $F_{k_y,j,k_y^\prime,j^\prime}=[1+\cos(\theta_{k_y,j}-\theta_{k_y^\prime,{j^\prime}})]/2$
is the form factor coming from the overlap of the single-particle nanoribbon wave functions, with
$\theta_{k_y,j} = \tan^{-1}(k_y/k_j)$, $V_{\text{e-h}}(q)$ is the effective electron-hole pairing interaction, and
 $E_j(k_y) =[(\epsilon_j(k_y) -\mu)^2+\Delta_j(k_y) ^2]^{1/2}$
 is the single-particle energy dispersion in the superfluid state
  for subband $j$.  The wave-vector $k_y^\prime$ is bounded by
the Brillouin zone boundary $\pm k_c$.  We truncate the sum over the subband index  at $j^\prime=j_c$, where $j_c$ is the lowest subband with a minimum above the graphene nanoribbon work function energy, taken  to be $\sim 4.5$ eV.
The chemical potential $\mu$ is fixed by the density equation,
\begin{eqnarray}
n &=& \frac{2}{W L_y}\sum_{j}^{j_c}\sum_{k_y}^{k_c} v_j(k_y)^2\ ;
\label{gapeqn2}\\
\label{vk2}
v_j(k_y)^2 &=& \frac{1}{2}\left[1 - \frac{\epsilon_j(k_y)-\mu }{E_j(k_y) }\right]\ .
\end{eqnarray}
The  factor $2$ in Eq.\ (\ref{gapeqn2}) is due to the spin degeneracy.

In the calculations we neglect screening, an
approximation that is justified
{\it a posteriori} as follows.  The screening is expected to be weak when the superfluidity lies
in the strongly-coupled BEC or BCS-BEC crossover regimes because  the superfluid gap in these regimes of pairing is  comparable to the  Fermi energy, resulting in a
large smearing of the Fermi surface, and compact electron-hole pairs  compared with their average spacing.  This makes their  mutual  interactions
 dipolar and weak.   In the BEC or BCS-BEC crossover
regimes we can thus expect screening to be weak, and so we  neglect it.  However for weak-coupled BCS superfluidity, screening
is known to be a strong effect,  which would make our  unscreened approximation a poor one in the BCS regime.  These effects have been documented  in the related system of
superfluidity in coupled electron-hole sheets with parabolic energy band
dispersion \cite{Perali2013,Maezono2013}.

Thus in Eq.\ (\ref{gapeqn}) we approximate the  electron-hole  pairing attraction   $V_{\text{e-h}}(q) $ by the bare Coulomb interaction.  For electrons and holes confined in nanoribbons of width $W$, separated by an insulating h-BN barrier of thickness $d$ and dielectric constant  $\kappa=3$, we obtain \cite{Brey2007},
\begin{equation}
V_{\text{e-h}}(q) = \frac{-2e^2}{\kappa W^2}\int_{0}^W \!\!\!\! \int_{0}^W\!\!\!\!\! dx^\prime dx K_0(|q|\sqrt{(x-x^\prime)^2+d^2}).
\label{Vunscr}
\end{equation}
Following Fig.\ 1 of Ref. \onlinecite{Brey2007}, we expect that  interband matrix elements will be small compared with  intraband matrix elements.  Thus in  Eq.\ (\ref{gapeqn}) we neglect the crosspairing terms,  where Cooper pairs would form with carriers
from different subbands.

\section*{Results}
\begin{figure}[t]
\centering
\includegraphics[width=15cm]{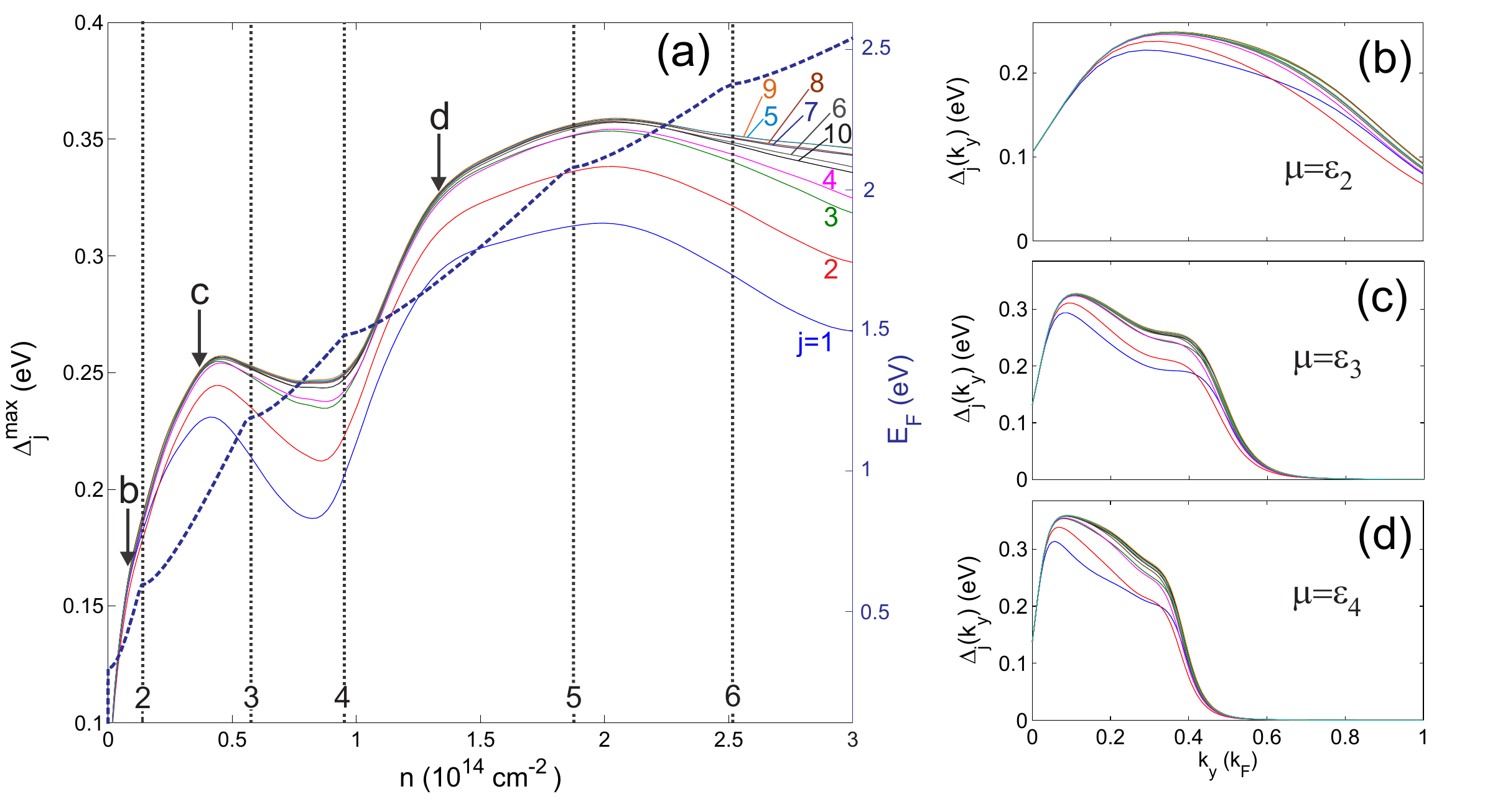}
\caption{
(a) Maximum superfluid gap $\Delta^{\mathrm{max}}_j$ for subbands $j=1,2\dots$ as  function of density $n$.  The dotted line shows the total Fermi energy $E_F$.
Nanoribbon
widths $W=2$ nm and barrier thickness $d=5$ nm. The densities at which $E_F$ enters the bottom of a new subband are indicated by the vertical lines.  Note that  the $\Delta^{\mathrm{max}}_j$ are all of order $E_F$.   Right panels (b) - (d): momentum-dependent gaps $\Delta_j(k_y)$ for subbands $j$ at densities marked by the arrows in panel (a) (at which $\mu=\varepsilon_{2},\varepsilon_3,\varepsilon_4$).}
\label{Deltasubband}
\end{figure}
Figure \ref{Deltamaxvsd} shows the maximum superfluid gap $\overline{\Delta}\,^{\mathrm{max}}$ as a function of the density $n$, averaged over the multiple subbands of the nanoribbons, calculated using Eqs.\ \ref{gapeqn} to \ref{Vunscr}.
$\overline{\Delta}\,^{\mathrm{max}}$ is the maximum value of the wave-vector
dependent $\overline{\Delta}(k_y)$ averaged with respect to the subband
index $j$.
The nanoribbon width is $W=2$ nm, and $d$ is the thickness of the insulating barrier.
The densities at which the Fermi energy enters the bottom of a new subband
$\epsilon_j$, $j=1,2,\dots$, are indicated by the vertical lines.

In Fig.\ \ref{Deltamaxvsd} we notice a local boost in
$\overline{\Delta}\,^{\mathrm{max}}$ near the minimum of each subband for barrier
thickness $d=5$ nm.  This boost
arises from shape resonance effects associated with the van Hove singularities
(Fig.\ \ref{DOS}(b)) and the quantum size effects in the pairing interaction.
However, for thinner barriers $d\alt 4$ nm, where the  electron-hole pairing becomes progressively stronger,  the shape resonance effects are masked by a mixing of the subbands caused by the large superfluid gap.
As $\overline{\Delta}\,^{\mathrm{max}}$ grows larger than the
typical spacing between subbands, the system becomes decreasing less one-dimensional
in character, thanks to the many channels available both  for Cooper pairing and for
forming the  superfluid condensate.

 \begin{figure}[h]
\centering
\includegraphics[width=8cm]{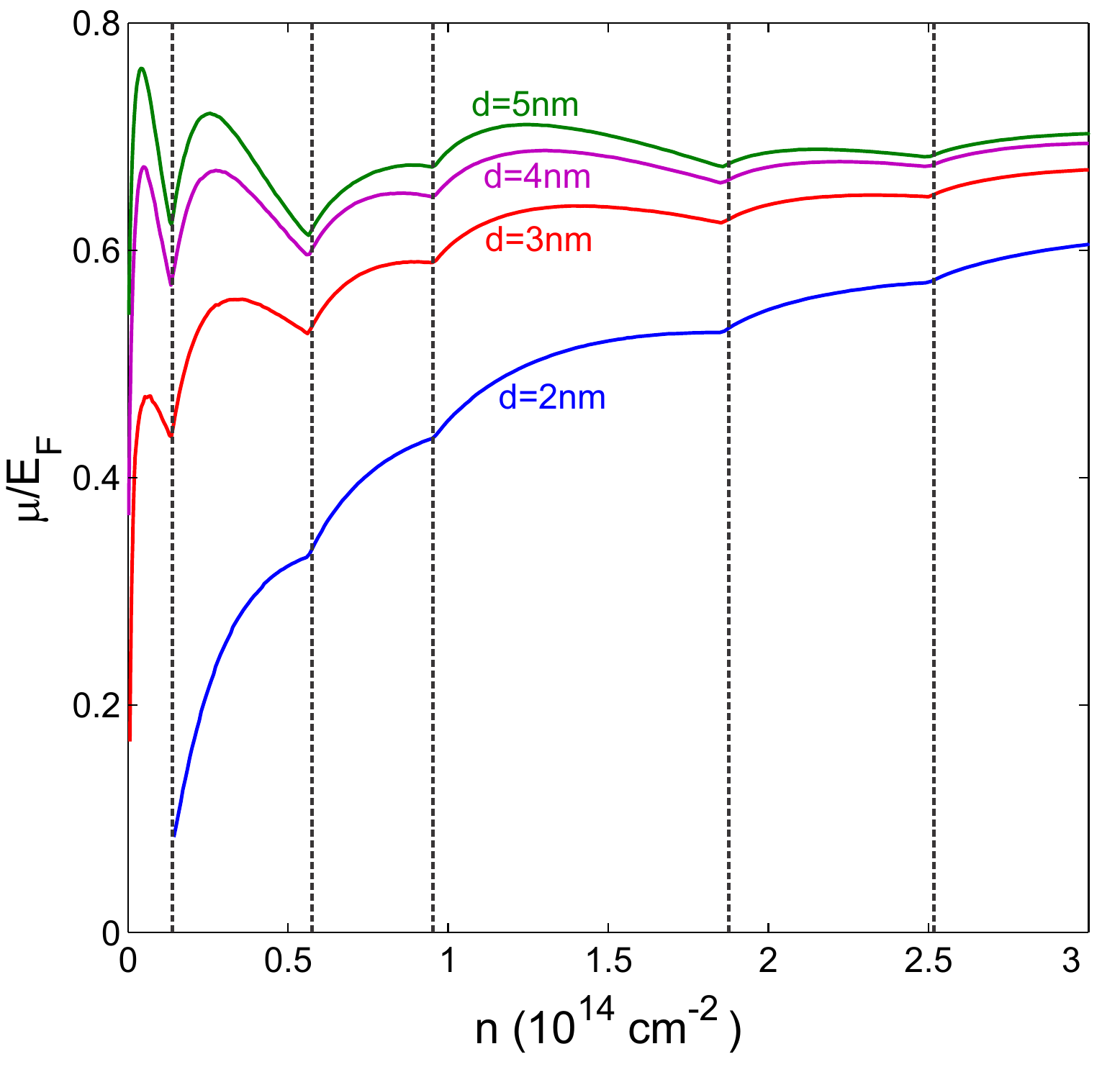}
\caption{Ratio of chemical potential $\mu$ to Fermi energy $E_F$ as function of density $n$.  $d$ is thickness of the insulating barrier separating the nanoribbons.  Nanoribbon width $W=2$ nm.  The vertical lines show the densities at which $E_F$ enters the bottom of a  subband.}
 \label{muoveref}
 \end{figure}
\begin{figure}[h]
\centering
\includegraphics[width=8cm]{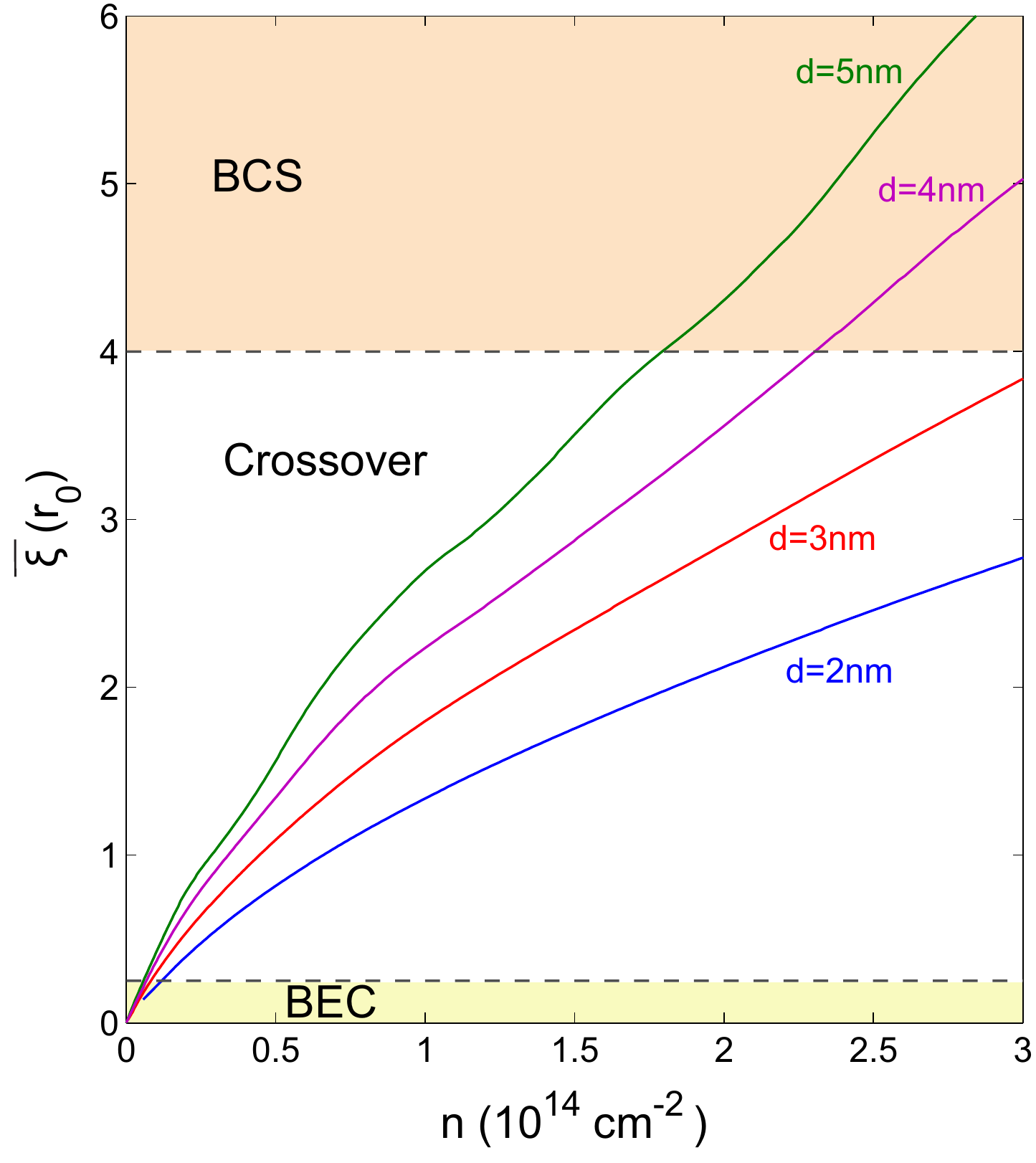}
\caption{
Pair correlation length $\overline{\xi}/r_0$ averaged over subbands as function of $n$, the carrier density.  $d$ is  thickness of the insulating barrier separating the nanoribbons.  Nanoribbon width $W=2$ nm.  }
 \label{xi}
 \end{figure}
Figure \ref{Deltasubband}(a) shows the maximum superfluid
gap $\Delta^{\texttt{max}}_j$ for the separate subbands $j$ as a
function of  $n$ for nanoribbon width  $W=2$ nm and barrier
thickness $d=5$ nm.
For comparison, the total Fermi energy $E_F$ of the non-interacting nanoribbon system at density $n$  is also shown.
The vertical lines  mark the densities at which
$E_F$ enters a new subband.
For the lowest conduction subbands, there is a notable local boost in
$\Delta^{\texttt{max}}_j$  as $E_F$ enters a new subband.
This boost takes the form of a shape resonance in the superfluid gaps
associated with a particular subband.
However even
for the lowest subbands, we lose some  fine structure of the shape
resonances because of mixing  by the gap of close lying subbands.
Over the density range shown, the $\Delta^{\texttt{max}}_j$ remain always of order $E_F$, and hence they lie in the strongly coupled regime.

Figures \ref{Deltasubband}(b)-(d), show the
momentum-dependence of the subband gaps $\Delta_j(k_y)$ for
densities at which the chemical potential
$\mu$ enters a new low-lying  subband (marked  in Fig.\   \ref{Deltasubband}(a) by the vertical arrows).
The peaks in $\Delta_j(k_y)$ are broad on the scale of $k_F=\pi/(2 r_0)$, the inverse of the average interparticle spacing, which
confirms that we are in the BCS-BEC crossover regime of compact electron-hole pairs.
In panels (c) and (d) of Fig.\ \ref{Deltasubband}, the multiple peaks of $\Delta_j(k_y)$ are associated with the different Fermi energies of the subbands $(k_F)_j$, displaying a remaining fermionic character of the Cooper pairing in the BCS-BEC crossover regime.

Figure \ref{muoveref} shows the chemical potential $\mu$
as a function of density $n$  for
nanoribbon width $W=2$ nm and  barrier thickness $d$.   $\mu$ is normalized to the
corresponding Fermi energy of the non-interacting nanoribbon system at density $n$.
The chemical potential is strongly renormalized with respect to
the Fermi energy over the full  range of $n$ and $d$ shown.   When $E_F$ enters a subband,
$\mu$ has a dip.  This is in contrast to the peak seen in the superfluid gap, and it is a shape antiresonance
caused by the shape-resonance-generated peak in the gap.
As $d$ is decreased and the pairing strength weakens, $\mu$ increases towards  $E_F$ and the shape (anti)resonances become sizeable, indicating
that the system has entered the BCS-BEC crossover regime.
In the case of $d=2$nm,  $\mu\ll E_F$ and the shape (anti)resonances are completely  smoothed out.  This is a result
of  large superfluid gaps and it signals that the system is in the  strong pairing BEC regime.
When the density increases, the system always
evolves towards the weaker pairing BCS regime for all values of $d$, with $\mu$ eventually arriving at $E_F$.

The average pair size of the Cooper pairs $\xi_j$ in
subband $j$ is defined as the expectation value of the square of the
relative coordinate of the Cooper pairs with respect to the square of the
BCS wave function projected in the subband.  This definition was originally
introduced in Ref.\ \onlinecite{Pistolesi1994} to investigate the different regimes of pairing in high-$T_c$
superconductivity in cuprates as a function of density.  It has been extended to a multigap superconductor
throughout the BCS-BEC crossover in Ref.\  \onlinecite{Guidini2014} and to a multigap quasi-one-dimensional superfluid of
ultracold fermions confined in cigar-shaped traps \cite{Shanenko2012}.
In  wave-vector space,
\begin{equation}
\label{csi_pair_eq}
\xi_j=\left[ \frac{\sum_{k_y} |\nabla _{k_y} (u_j(k_y)v_j(k_y))|^2}{\sum_{k_y} (u_j(k_y)v_i(k_y))^2}\right]^{\frac{1}{2}}\ ,
\end{equation}
where $u_j(k_y)^2=1-v_j(k_y)^2$.

Figure \ref{xi} shows the pair correlation length $\overline{\xi}/r_0$ as a function of density.  $\overline{\xi}/r_0$ is the partial average pair size for each subband averaged over the subbands.    The nanoribbon width
$W=2$ nm. We designate $\overline{\xi}/r_0<0.25$ as the BEC regime,  $0.25<\overline{\xi}/r_0<4$ the BCS-BEC crossover regime, and  $\overline{\xi}/r_0>4$ the BCS regime.
As discussed, our approximation of neglecting screening is expected to be a good one for  densities lying in the BCS-BEC crossover and BEC regimes.
As expected, the density range for the strongly-coupled
regime contracts with increasing barrier thickness $d$ because the pairing becomes weaker.

We have neglected effects from impurities and disorder.  We expect these effects to be small since while there is no
direct information on impurity and disorder effects in graphene nanoribbons,
but based on properties of analogous  coupled
electron-hole graphene monolayers, charge impurities concentrations up to
$n_i < k_F/(\pi d)$ are not expected  to destroy superfluidity  \cite{Bistritzer}.
At graphene-hBN interfaces, the charge impurity density
$n_i\agt 10^{10}$ cm$^{-2}$ \cite{Martin}, so for  $d \alt 5$ nm, the inequality is satisfied provided $n \agt 3\times 10^6$ cm$^{-2}$.  This density  is orders of
magnitude less than current experimental densities.

\section*{Conclusions}

The superfluid gaps in our coupled electron-hole nanoribbon systems
are large  in absolute value and comparable to the Fermi energy.  The quasi-one-dimensional confinement and the superfluid shape resonances due to quantum size
effects both play an important role here.  The van Hove singularities in the densities of states act
non-linearly through the gap
equation to significantly enhance the magnitude of the superfluid gaps.
In the range of nanoribbon densities and barrier separations considered, we find that the electron-hole superfluid is for the most part in the strongly coupled pairing regime, and so Coulomb screening effects are expected to be weak.

When the superfluid gaps are comparable to the subband energy separations, the gaps
mix the subbands and this results in a rounding of the shape resonances. This effect is most pronounced for small separations between the nanoribbons where  the  electron-hole coupling is particularly strong.
For larger separations, the electron-hole coupling is weaker and the superfluid gaps are smaller.  This results in weaker
subband mixing.  When this is the case, the shape resonances are sharper which strengthens the
local amplification of the gaps.

In our quasi-one-dimensional system there is no direct
link between the superfluid transition temperature and
the size of the superfluid gaps calculated within mean field.
In our proposed device the zero temperature superfluid gaps are comparable
to the Fermi energy, and can be of order of hundreds of meV.
Thus high transition temperature electron-hole superfluidity
could be expected, with  properties that are tuneable by changing the density.  The device configurations we propose are experimentally realizable with currently available technologies.  A superlattice formed of such nanoribbon devices
could further stabilize the electron-hole superfluid phase over  large areas.

\section*{Acknowledgements}

M.Z. acknowledges support by the Flemish Science
Foundation (FWO-Vl), the University Research Fund (BOF),
and the European Science Foundation (POLATOM).
A.P. and D.N. acknowledge support by the University of Camerino
FAR project CESEMN.
The authors thank the colleagues involved in the MultiSuper
International Network (http://www.multisuper.org) for exchange of ideas and
suggestions for this work.

\end{document}